\documentclass[preprint]{revtex4-1}
\usepackage{amssymb,graphicx,amssymb}

\makeatletter

\@ifundefined{textcolor}{}
{%
 \definecolor{BLACK}{gray}{0}
 \definecolor{WHITE}{gray}{1}
 \definecolor{RED}{rgb}{1,0,0}
 \definecolor{GREEN}{rgb}{0,1,0}
 \definecolor{BLUE}{rgb}{0,0,1}
 \definecolor{CYAN}{cmyk}{1,0,0,0}
 \definecolor{MAGENTA}{cmyk}{0,1,0,0}
 \definecolor{YELLOW}{cmyk}{0,0,1,0}
}

\makeatother

\begin{document}
\pagestyle{plain}

\title{The Dark Matter Self-Interaction and Its Impact on the Critical Mass
for Dark Matter Evaporations Inside the Sun}

\author{Chian-Shu Chen$^{*,\ddagger}$\footnote{chianshu@phys.sinica.edu.tw},
Fei-Fan Lee$^{\dagger}$\footnote{fflee@mail.nctu.edu.tw}, Guey-Lin
Lin$^{\dagger}$\footnote{glin@cc.nctu.edu.tw} and Yen-Hsun Lin$^{\dagger}$\footnote{chris.py99g@g2.nctu.edu.tw}}

\affiliation{\emph{$^{*}$}Physics Division, National Center for Theoretical Science,
Hsinchu 30010, Taiwan\\
$^{\dagger}$Institute of Physics, National Chiao Tung University,
Hsinchu 30010, Taiwan\\
$^{\ddagger}$Department of Physics, National Tsing Hua University,
Hsinchu 30010, Taiwan}
\begin{abstract}
We study the capture, annihilation and evaporation of dark matter
(DM) inside the Sun. It has been shown that the DM self-interaction
can increase the DM number inside the Sun. We demonstrate that this
enhancement becomes more significant in the regime of small DM mass,
given a fixed DM self-interaction cross section. This leads to the
enhancement of neutrino flux from DM annihilation. On the other hand,
for DM mass as low as as a few GeVs, not only the DM-nuclei scatterings
can cause the DM evaporation, DM self-interaction also provides non-negligible
contributions to this effect. Consequently, the critical mass for
DM evaporation (typically $3 \sim 4$ GeV without the DM self-interaction)
can be slightly increased. We discuss the prospect of detecting DM
self-interaction in IceCube- PINGU using the annihilation channels
$\chi\chi\rightarrow\nu\bar{\nu},\:\tau^{-}\tau^{+}$ as examples.
The PINGU sensitivities to DM self-interaction cross section $\sigma_{\chi\chi}$
are estimated for track and cascade events.
\end{abstract}
\maketitle

\section{Introduction}

In this talk, we present the general framework of DM capture, annihilation
and evaporation in the Sun. The capture of galactic DM by the Sun
through DM-nuclei collisions was first proposed and calculated in
Refs.~\cite{Steigman:1997vs,Press:1985ug,Faulkner:1985rm,Griest:1986yu}.
It was then observed that the assumption of DM thermal distribution
according to the average temperature of the Sun is a good approximation
for capture and annihilation processes, but the correction to the
evaporation mass can reach to 8\% in the true distribution calculation~\cite{Gould:1987ju}.
The abundance of DM inside the Sun hence results from the balancing
among DM capture, annihilation and evaporation processes.

In our study, we note that for both collisionless cold DM and warm
DM there exists a so-called core/cusp problem~\cite{deBlok:2009sp}
which addresses the discrepancy between the computational structure
simulation and the actual observation~\cite{Moore:1994yx,Flores:1994gz,Navarro:1996gj}.
DM self-interaction has been introduced to resolve this inconsistency~\cite{Spergel:1999mh}.
Constraints on the ratio of DM self-interaction cross section to the
DM mass, $0.1<\sigma_{\chi\chi}/m_{\chi}<1.0~({\rm {cm}^{2}/g)}$,
were obtained from observations of various galactic structures~\cite{Randall:2007ph,Rocha:2012jg,Peter:2012jh,Zavala:2012us}.
We note that the authors in Ref.~\cite{Albuquerque:2013xna} used
the IceCube data~\cite{Aartsen:2012kia} to constrain the magnitude
of $\sigma_{\chi\chi}$ for $m_{\chi}$ in the range of ${\cal O}(10)$~GeV
to ${\cal O}(1)$~TeV (also see the study of high energy neutrino
flux from DM annihilation within the Sun with the inclusion of DM
self-interaction in Ref.~\cite{Zentner2009}). In their work the
evaporation effect can be neglected for the considered DM mass range.
In this presentation we shall concentrate on the low mass region of
${\cal O}(1)$~GeV DM mass since such a mass range has not been probed
by the IceCube data mentioned above. Furthermore, this is also the
mass range where the indirect search is crucial. In the case of spin-independent
interaction, the sensitivity of DM direct search quickly turns poor
for $m_{\chi}$ less than $10$ GeV \cite{Cushman:2013zza}. Therefore
the IceCube-PINGU~\cite{Aartsen2014} detector with an $1$ GeV threshold
energy could be more sensitive than some of the direct detection experiments
for $m_{\chi}<10$ GeV. For spin-dependent interaction, the IceCube-PINGU
sensitivity has been estimated to be much better than constraints
set by direct detection experiments~\cite{Feng:2014uja}.

\section{DM accumulation in the Sun}

\begin{figure*}[t]
\begin{centering}
\includegraphics[width=0.46\textwidth]{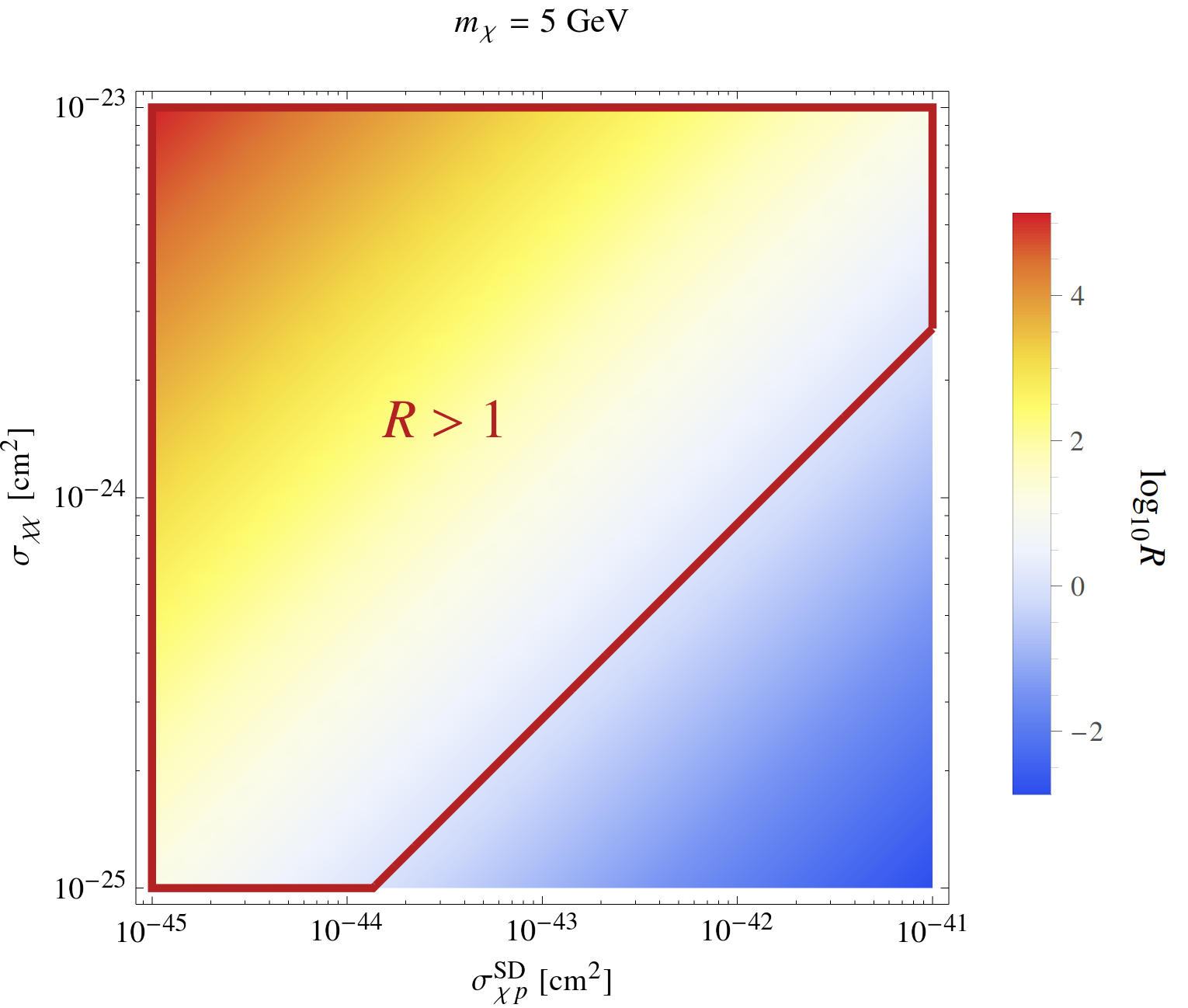}\ \includegraphics[width=0.46\textwidth]{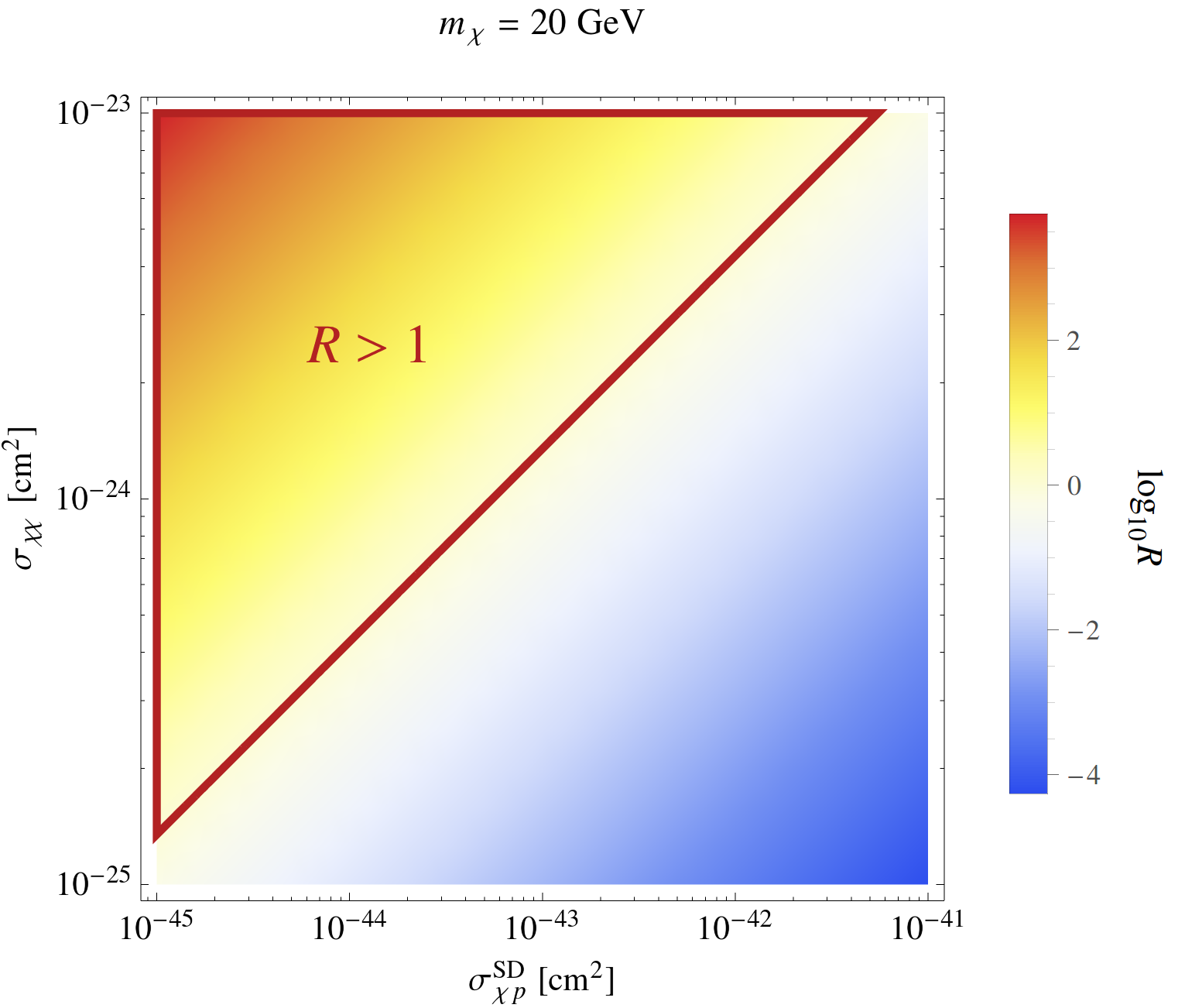}
\par\end{centering}

\begin{centering}
\includegraphics[width=0.46\textwidth]{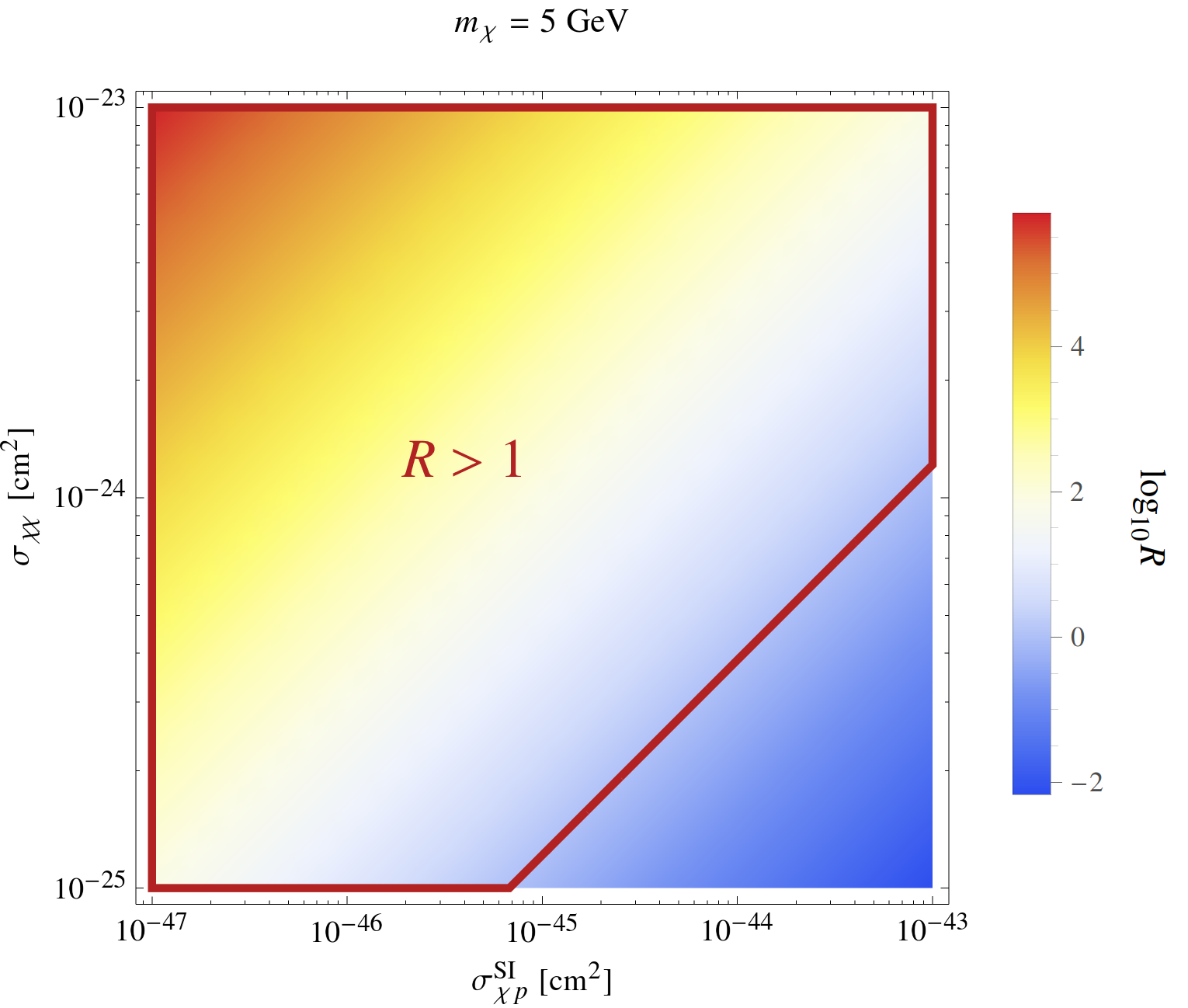}\enskip{}\includegraphics[width=0.46\textwidth]{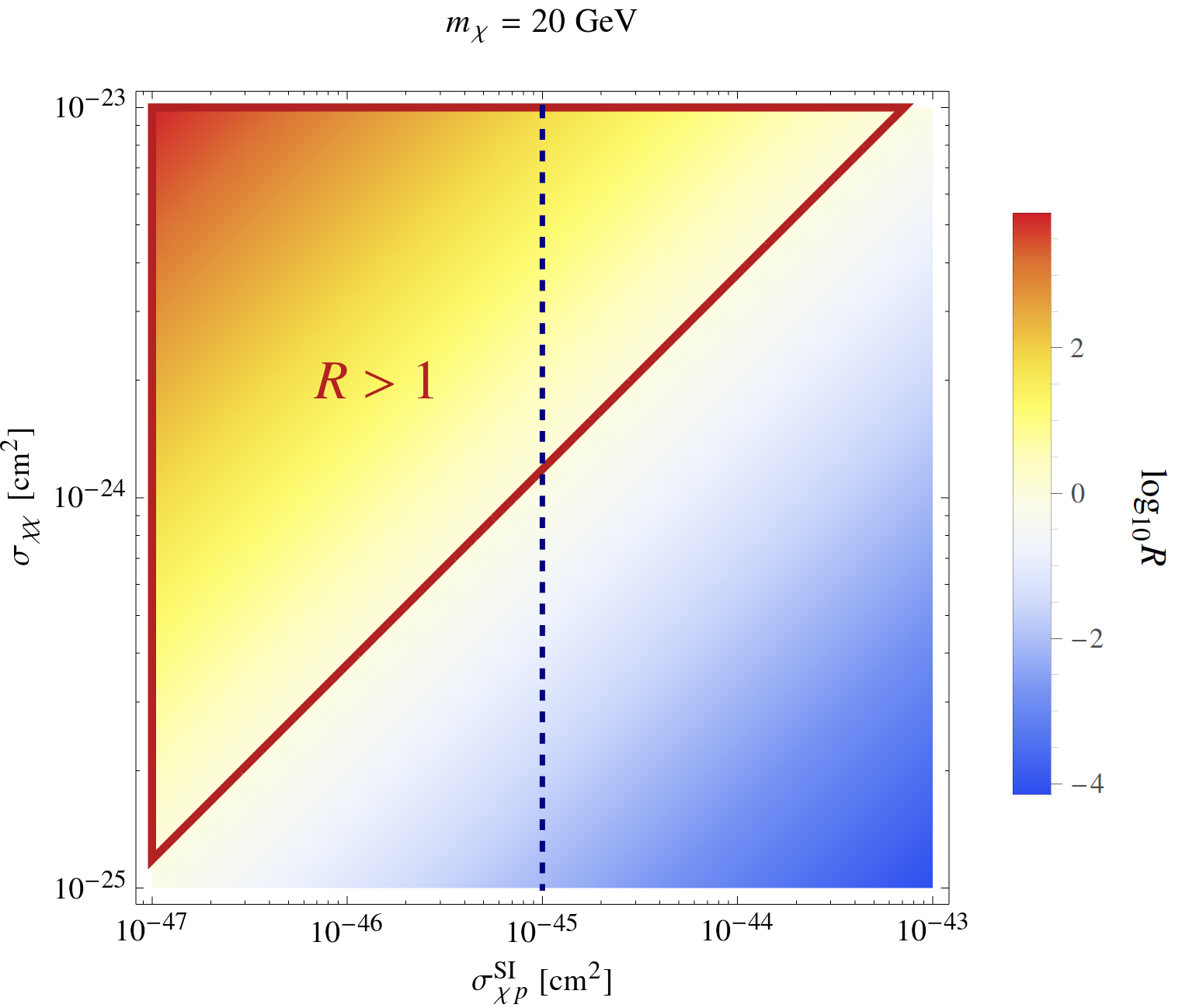}
\par\end{centering}

\protect\caption{\label{fig:Rs_2D}Ratio $R$ over the $\sigma_{\chi p}-\sigma_{\chi\chi}$
plane. The upper panel is for SD interaction and the lower panel is
for SI interaction. The red-circled region is for $R>1$. The region
to the right of the blue-dashed line is excluded by LUX.}
\end{figure*}
The evolution of DM particles captured by the solar gravity is described
by the following differential equation, 
\begin{equation}
\frac{dN_{\chi}}{dt}=C_{c}+(C_{s}-C_{e})N_{\chi}-(C_{a}+C_{se})N_{\chi}^{2}\label{eq:evo_eq}
\end{equation}
with $N_{\chi}$ the DM number in the Sun, and $C_{c}$ the rate at
which DM are captured by the Sun. One has \cite{Bertone2005} 
\begin{equation}
C_{c}^{{\rm SD}}\simeq3.35\times10^{24}\textrm{ s}^{-1}\left(\frac{\rho_{0}}{0.3\textrm{ GeV/cm}^{3}}\right)\left(\frac{270\textrm{ km/s}}{\bar{v}}\right)^{3}\times\left(\frac{\textrm{GeV}}{m_{\chi}}\right)^{2}\left(\frac{\sigma_{{\rm H}}^{{\rm SD}}}{10^{-6}\textrm{ pb}}\right)\label{eq:capture_SD}
\end{equation}
for spin-dependent (SD) interaction; 
\begin{equation}
C_{c}^{{\rm SI}}\simeq1.24\times10^{24}\textrm{ s}^{-1}\left(\frac{\rho_{0}}{0.3\textrm{ GeV/cm}^{3}}\right)\times\left(\frac{270\textrm{ km/s}}{\bar{v}}\right)^{3}\left(\frac{\textrm{GeV}}{m_{\chi}}\right)^{2}\left(\frac{2.6\sigma_{{\rm H}}^{{\rm SI}}+0.175\sigma_{{\rm He}}^{{\rm SI}}}{10^{-6}\textrm{ pb}}\right)
\end{equation}
for spin-independent (SI) interaction. Here $\rho_{0}$ is the local
DM density, $\bar{v}$ is the DM velocity dispersion and $\sigma_{A}$
is DM-nuclei cross section for SD or SI interaction.

$C_{s}$ is the rate at which DM are captured due to their scattering
with DM that have already been trapped in the Sun \cite{Zentner2009},
\begin{equation}
C_{s}=\sqrt{\frac{3}{2}}n_{\chi}\sigma_{\chi\chi}v_{{\rm esc}}(R_{\odot})\frac{v_{{\rm esc}}(R_{\odot})}{\overline{v}}\left\langle \hat{\phi}_{\chi}\right\rangle \frac{\textrm{erf}(\eta)}{\eta}
\end{equation}
where $v_{{\rm esc}}(R_{\odot})$ is the solar escape at the surface
and $\eta^{2}=3(v_{\odot}/\bar{v})^{2}/2$ with $v_{\odot}$ the velocity
of the Sun.

$C_{e}$ is the the DM evaporation rate due to DM-nuclei interactions
\cite{Busoni2013},

\begin{equation}
C_{e}\simeq\frac{8}{\pi^{3}}\sqrt{\frac{2m_{\chi}}{\pi T_{\chi}(\bar{r})}}\frac{v_{\textrm{esc}}^{2}(0)}{\bar{r}^{3}}\times\exp\left(-\frac{m_{\chi}v_{\textrm{esc}}^{2}(0)}{2T_{\chi}(\bar{r})}\right)\Sigma_{{\rm evap}},\label{eq:evap_rate}
\end{equation}
where $v_{{\rm esc}}(0)$ is the solar escape velocity at the core,
$T_{\chi}$ is the DM temperature in the Sun, and $\bar{r}$ is average
DM orbit radius. The quantity $\Sigma_{{\rm evap}}$ is the sum of
the scattering cross section of all the nuclei within a radius $r_{95\%}$,
where the solar temperature has dropped to 95\% of the DM temperature.

$C_{a}$ is the DM annihilation rate given by,
\begin{equation}
C_{a}\simeq\frac{\left\langle \sigma v\right\rangle V_{2}}{V_{1}^{2}},
\end{equation}
where 
\begin{equation}
V_{j}\simeq6.5\times10^{28}\textrm{ cm}^{3}\left(\frac{10\textrm{ GeV}}{jm_{\chi}}\right)^{3/2}.
\end{equation}

$C_{se}$ is the evaporation rate induced by the interaction between
DM particles in the Sun given by \cite{Chen2014},

\begin{equation}
C_{se}=\frac{\int_{\odot}\frac{dC_{se}}{dV}d^{3}r}{\left(\int_{\odot}n_{\chi}(r)d^{3}r\right)^{2}},
\end{equation}
where
\begin{equation}
\frac{dC_{se}}{dV}=\frac{4}{\sqrt{\pi}}\sqrt{\frac{m_{\chi}}{2T_{\chi}}}\frac{n_{0}^{2}\sigma_{\chi\chi}}{m_{\chi}}\exp\left[-\frac{2m_{\chi}\phi(r)}{T_{\chi}}\right]\times\exp\left[-\frac{E_{{\rm esc}}(r)}{T_{\chi}}\right]\tilde{K}(m_{\chi})
\end{equation}
and
\begin{equation}
n_{\chi}(r)=n_{0}\exp\left(-\frac{m_{\chi}\phi(r)}{T_{\chi}}\right).
\end{equation}
Here $n_{0}$ is the DM number in the solar core, $\phi$ is the solar
gravitational potential, $E_{{\rm esc}}(r)$ is the escape energy
at radius $r$ inside the Sun and $\tilde{K}(m_{\chi})$ is defined
in the appendix of Ref.~\cite{Chen2014}. All the coefficients $C_{c,a,e,s,se}$
are positive and time-independent.

With $N_{\chi}(0)=0$ as the initial condition, the general solution
to Eq.~(\ref{eq:evo_eq}) is 
\begin{equation}
N_{\chi}(t)=\frac{C_{c}\tanh(t/\tau_{A})}{\tau_{A}^{-1}-(C_{s}-C_{e})\tanh(t/\tau_{A})/2},
\end{equation}
with 
\begin{equation}
\tau_{A}=\frac{1}{\sqrt{C_{c}(C_{a}+C_{se})+(C_{s}-C_{e})^{2}/4}}
\end{equation}
the time-scale for the DM number in the Sun to reach the equilibrium.
If the equilibrium state is achieved, i.e., $\tanh(t/\tau_{A})\sim1$,
one has 
\begin{equation}
N_{\chi,{\rm eq}}=\sqrt{\frac{C_{c}}{C_{a}+C_{se}}}\left(\pm\sqrt{\frac{R}{4}}+\sqrt{\frac{R}{4}+1}\right),\label{eq:n_chi_equ}
\end{equation}
where one takes the positive sign for $C_{s}>C_{e}$ and the negative
sign for $C_{e}>C_{s}$. The dimensionless parameter $R$ is defined
as 
\begin{equation}
R\equiv\frac{(C_{s}-C_{e})^{2}}{C_{c}(C_{a}+C_{se})}.
\end{equation}
This ratio determines whether the self-interaction is important ($R>1$)
or not ($R<1$). The region for $R>1$ is shown in Fig.~\ref{fig:Rs_2D}.
Note that the region for $R>1$ shrinks when $m_{\chi}$ becomes heavier.
It implies that the self-interaction is significant for lighter DM.

\begin{figure*}
\begin{centering}
\includegraphics[width=0.44\textwidth]{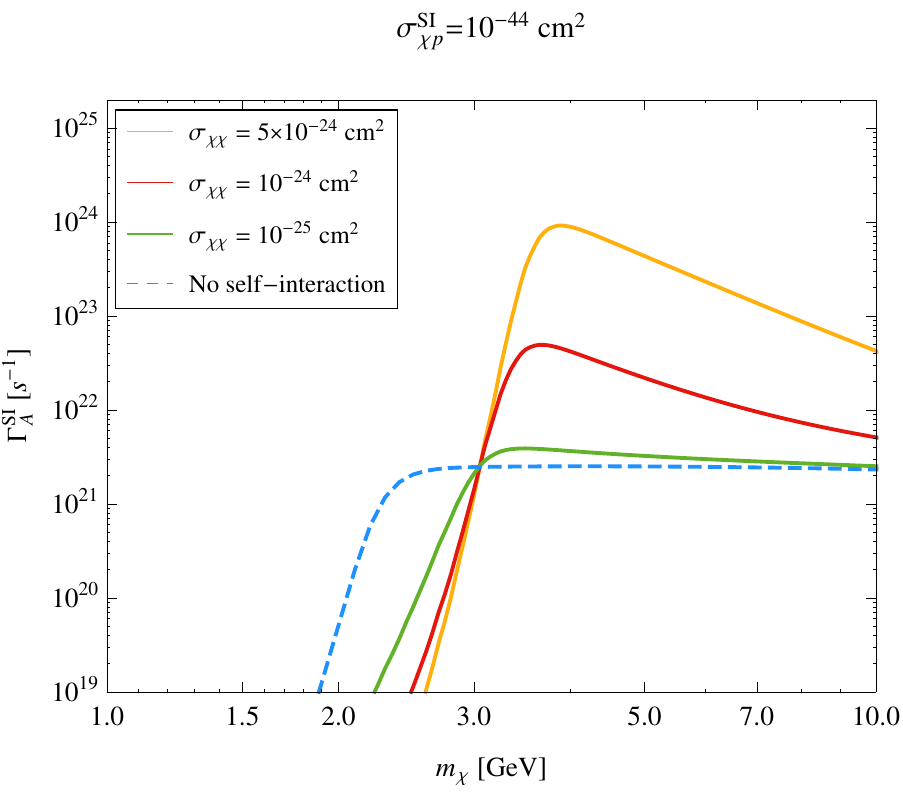}\enskip{}\includegraphics[width=0.44\textwidth]{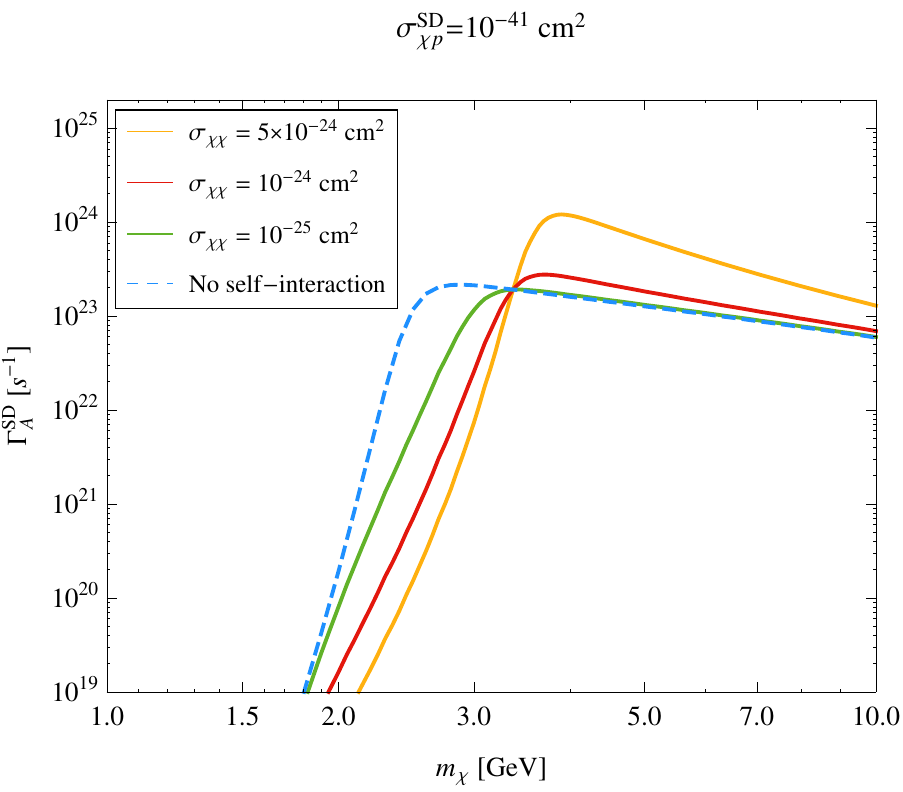}
\par\end{centering}

\protect\caption{\label{fig:WIMP_number_ann}The annihilation rate $\Gamma_{A}$ of
the captured DM inside the Sun. The left one assumes DM-nuclei scattering
is dominated by SI interaction while the right one assumes such scattering
is dominated by SD interaction.}
\end{figure*}

Thus, the DM total annihilation rate in the Sun's core is given by
\begin{equation}
\Gamma_{A}=\frac{C_{a}}{2}N_{\chi,{\rm eq}}^{2}=\frac{1}{2}\frac{C_{c}C_{a}}{C_{a}+C_{se}}\left(\pm\sqrt{\frac{R}{4}}+\sqrt{\frac{R}{4}+1}\right)^{2}.\label{annihilation}
\end{equation}
Where the sign convention is identical to that Eq.~(\ref{eq:n_chi_equ}).
The result for $\Gamma_{A}$ is shown in Fig.~\ref{fig:WIMP_number_ann}.

\section{Probing DM self-interaction at IceCube-PINGU}

\begin{figure*}[t]
\begin{centering}
\includegraphics[width=0.44\textwidth]{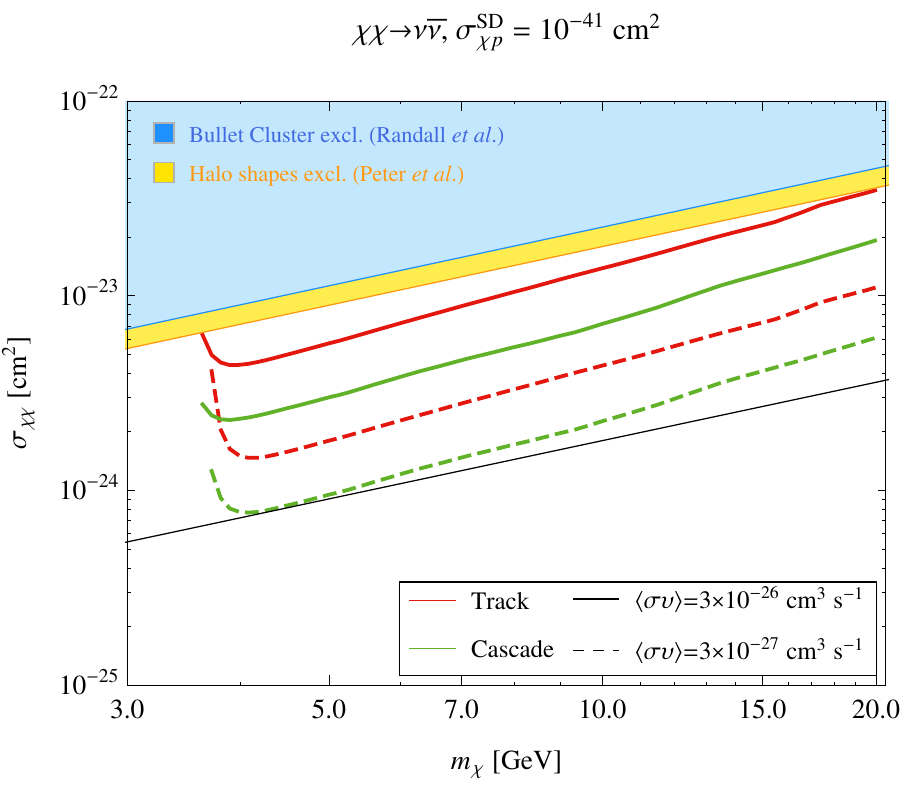}\enskip{}\includegraphics[width=0.44\textwidth]{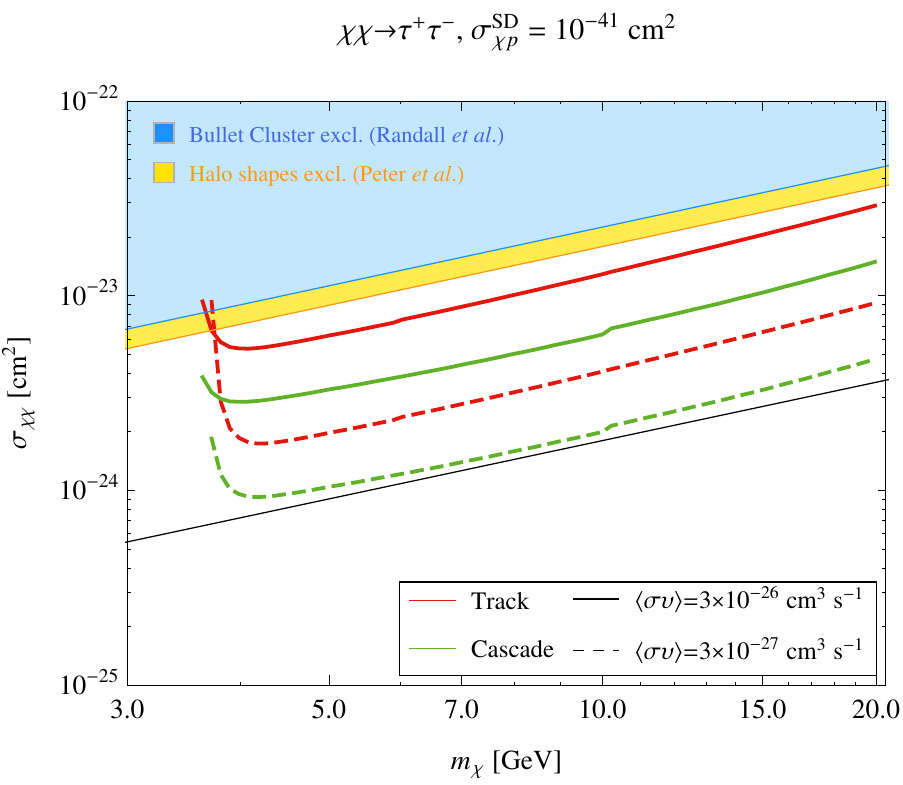}
\par\end{centering}

\begin{centering}
\includegraphics[width=0.44\textwidth]{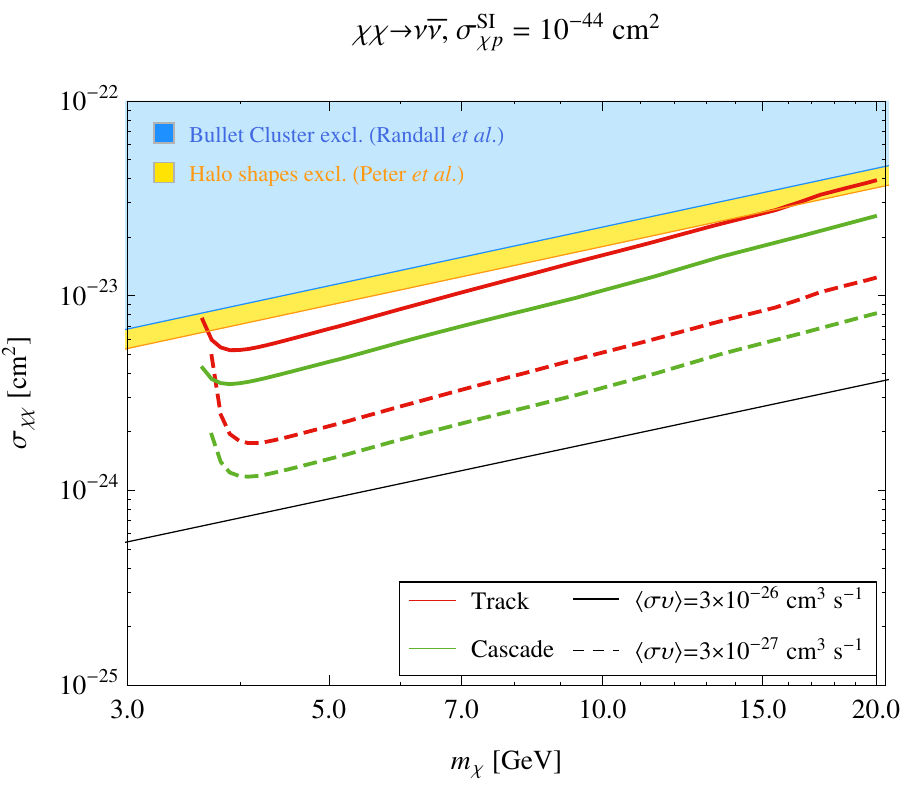}\enskip{}\includegraphics[width=0.44\textwidth]{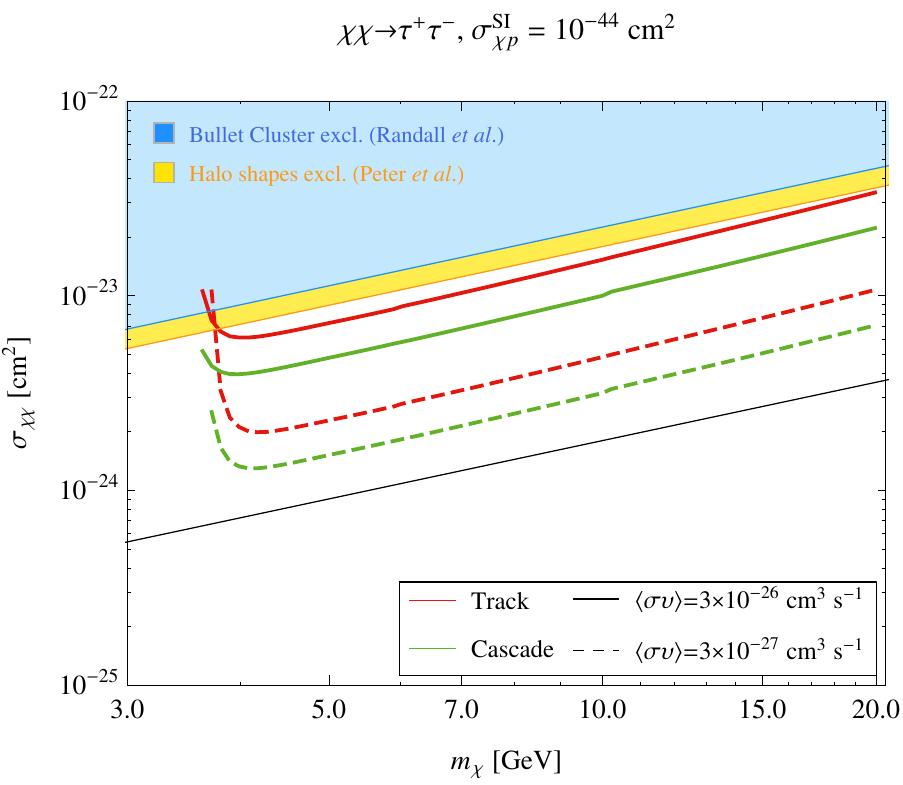}
\par\end{centering}

\protect\caption{\label{fig:Cons}The IceCube-PINGU sensitivities to DM self-interaction
cross section $\sigma_{\chi\chi}$ as a function of $m_{\chi}$. The
upper panel is the DM-nucleus interaction inside the Sun which is
assumed to be dominated by SD interaction. The lower panel is assumed
to be dominated by SI interaction.}
\end{figure*}
To probe DM self-interaction for small $m_{\chi}$, we consider DM
annihilation channels, $\chi\chi\rightarrow\tau^{+}\tau^{-}$ and
$\nu\bar{\nu}$, for producing neutrino final states to be detected
by IceCube-PINGU~\cite{Aartsen2014}. The neutrino differential flux
of flavor $i$, $\Phi_{\nu_{i}}$, from $\chi\chi\rightarrow f\bar{f}$
can be expressed as 
\begin{equation}
\frac{d\Phi_{\nu_{i}}}{dE_{\nu_{i}}}=P_{\nu_{j}\to\nu_{i}}(E_{\nu})\frac{\Gamma_{A}}{4\pi R_{\odot}^{2}}\sum_{f}B_{f}\left(\frac{dN_{\nu_{j}}}{dE_{\nu_{j}}}\right)_{f}\label{eq:neutrino_flux}
\end{equation}
where $R_{\odot}$ is the distance between the neutrino source and
the detector, $P_{\nu_{j}\to\nu_{i}}(E_{\nu})$ is the neutrino oscillation
probability during the propagation, $B_{f}$ is the branching ratio
corresponding to the channel $\chi\chi\rightarrow f\bar{f}$ , $dN_{\nu}/dE_{\nu}$
is the neutrino spectrum at the source, and $\Gamma_{A}$ is the DM
annihilation rate in the Sun. To compute $dN_{\nu}/dE_{\nu}$, we
employed \texttt{WimpSim} \cite{Blennow:2007tw} with a total of 50,000
Monte-Carlo generated events.

The neutrino event rate in the detector is given by 
\begin{equation}
N_{\nu}=\int_{E_{{\rm th}}}^{m_{\chi}}\frac{d\Phi_{\nu}}{dE_{\nu}}A_{\nu}(E_{\nu})dE_{\nu}d\Omega\label{eq:nu_event}
\end{equation}
where $E_{\textrm{th}}$ is the detector threshold energy, $d\Phi_{\nu}/dE_{\nu}$
is the neutrino flux from DM annihilation, $A_{\nu}$ is the detector
effective area, and $\Omega$ is the solid angle. We study both muon
track events and cascade events induced by neutrinos. The PINGU module
will be implanted inside the IceCube in the near future~\cite{Aartsen2014}
and can be used to probe neutrino energy down to ${\cal O}(1)$~GeVs.

The atmospheric background event rate can also be calculated by Eq.~(\ref{eq:nu_event})
with $d\Phi_{\nu}/dE_{\nu}$ replaced by the atmospheric neutrino
flux. Hence 
\begin{equation}
N_{{\rm atm}}=\int_{E_{{\rm th}}}^{E_{{\rm max}}}\frac{d\Phi_{\nu}^{{\rm atm}}}{dE_{\nu}}A_{\nu}(E_{\nu})dE_{\nu}d\Omega.\label{eq:atm_event}
\end{equation}
In our calculation, the atmospheric neutrino flux $d\Phi_{\nu}^{{\rm atm}}/dE_{\nu}$
is taken from Refs.~\cite{Aartsen:2012uu,Honda:2006qj}. We set $E_{{\rm max}}=m_{\chi}$
in order to compare with the DM signal.

The angular resolution for IceCube-PINGU detector at $E_{\nu}=5$
GeV is roughly $10^{\circ}$~\cite{Aartsen2014}. Hence we consider
neutrino events arriving from the solid angle range $\Delta\Omega=2\pi(1-\cos\psi)$
surrounding the Sun with $\psi=10^{\circ}$. We present the IceCube-PINGU
sensitivity to $\sigma_{\chi\chi}$ in the DM mass region $3~{\rm GeV}<m_{\chi}<20~{\rm GeV}$
for both SD and SI cases in Fig.~\ref{fig:Cons}. The sensitivities
to $\sigma_{\chi\chi}$ are taken to be $2\sigma$ significance for
5 years of data taking. The shadow areas in the figures represent
those parameter spaces disfavored by the Bullet Cluster and halo shape
analyses. Below the black solid line, the DM self-interaction is too
weak to resolve the core/cusp problem of the structure formation.
Two benchmark values of thermal average cross section, $\langle\sigma v\rangle=3\times10^{-26}~{\rm cm^{3}s^{-1}}~{\rm and}~\langle\sigma v\rangle=3\times10^{-27}~{\rm cm^{3}s^{-1}}$
are used for our studies. We note that the latter value for $\langle\sigma v\rangle$
does not contradict with the relic density, since DM annihilation
inside the Sun occurs much later than the period of freeze-out.

We take $\sigma_{\chi p}^{{\rm SD}}=10^{-41}~{\rm cm^{2}}$ for SD
interaction, and take $\sigma_{\chi p}^{{\rm SI}}=10^{-44}~{\rm cm^{2}}$
for SI interaction. We stress that $\sigma_{\chi p}^{{\rm SD}}=10^{-41}~{\rm cm^{2}}$
is below the lowest value of IceCube bound $\sigma_{\chi p}^{{\rm SD}}\sim10^{-40}~{\rm cm^{2}}$
at $m_{\chi}\sim300$ GeV~\cite{Aartsen:2012kia}. For SI interaction,
$\sigma_{\chi p}^{{\rm SI}}=10^{-44}~{\rm cm^{2}}$ is below the LUX
bound for $m_{\chi}<8$ GeV~\cite{LUX2013}. We find that cascade
events provide better sensitivities to DM self-interaction than track
events do in all cases. One can also see that the sensitivity to $\sigma_{\chi\chi}$
becomes better for smaller annihilation cross section $\langle\sigma v\rangle$
for a fixed $\sigma_{\chi p}$, as noted in earlier works~\cite{Zentner2009,Albuquerque:2013xna}
which neglect both $C_{e}$ and $C_{se}$. This is evident from Eq.~(\ref{annihilation})
since $R$ increases as $C_{a}$ decreases. It is instructive to take
the limit $R\gg1$ such that $\Gamma_{A}\to(C_{c}C_{a})R/2(C_{a}+C_{se})$
for $C_{s}>C_{e}$. It is easily seen that $\Gamma_{A}$ is inversely
proportional to $C_{a}$ (in the mass range that $C_{se}$ is negligible)
and is independent of $C_{c}$. In other words, only $C_{s}$ and
$C_{a}$ determine the annihilation rate (we are in the region that
$C_{e}$ is suppressed as compared to $C_{s}$). We also see that
the sensitivity to $\sigma_{\chi\chi}$ does become significantly
worse as $m_{\chi}\to4$ GeV. This is the critical $m_{\chi}$ below
which the DM evaporations from the Sun is important.

\section{Conclusion}

We have presented the time evolution of DM number trapped inside the
Sun with DM self-interaction considered. We focused on the low $m_{\chi}$
range which requires the consideration of evaporation effects due
to both DM-nuclei and DM-DM scatterings. The parameter region for
the trapped DM inside the Sun to reach the equilibrium state is presented.
We also found that the inclusion of DM self-interaction can increase
the number of trapped DM as well as raise the evaporation mass scale.
The parameter space on $\sigma_{\chi\chi}-\sigma_{\chi p}^{{\rm SD\,(SI)}}$
plane for significant enhancement on trapped DM number ($R>1$) is
identified. The parameter space for $R>1$ becomes larger for smaller
$m_{\chi}$. For $C_{s}<C_{e}$, the condition $R>1$ leads to the
suppression of neutrino flux, since the first term on the right hand
side of Eq.~(\ref{annihilation}) is negative. We have proposed to
study $\sigma_{\chi\chi}$ with the future IceCube-PINGU detector
where the energy threshold can be lowered down to $1$ GeV. We considered
cascade and track events resulting from neutrino flux induced by DM
annihilation channels $\chi\chi\to\nu\bar{\nu}$ and $\chi\chi\to\tau^{+}\tau^{-}$
inside the Sun. We found that cascade events always provide better
sensitivity to $\sigma_{\chi\chi}$. The sensitivity to $\sigma_{\chi\chi}$
is also improved with a smaller DM annihilation cross section $\langle\sigma v\rangle$.

\section*{Acknowledgement}

We thank S. Palomares-Ruiz for a very useful comment. CSC is supported
by the National Center for Theoretical Sciences, Taiwan; FFL, GLL,
and YHL are supported by Ministry of Science and Technology, Taiwan
under Grant No. 102-2112-M-009-017.


\begin{thebibliography}{10}
\bibitem{Steigman:1997vs} G.~Steigman, C.~L.~Sarazin, H.~Quintana
and J.~Faulkner, 
 Astron. J. 83 (1978) 1050-1061; Srednicki, M.A. (ed.): Particle physics
and cosmology 207-218.

\bibitem{Press:1985ug} D.~N.~Spergel and W.~H.~Press, 
 Astrophys.\ J.\ \textbf{294}, 663 (1985); 
W.~H.~Press and D.~N.~Spergel, 
 Astrophys.\ J.\ \textbf{296}, 679 (1985). 
 

\bibitem{Faulkner:1985rm} J.~Faulkner and R.~L.~Gilliland, 
 Astrophys.\ J.\ \textbf{299}, 994 (1985). 
 

\bibitem{Griest:1986yu} K.~Griest and D.~Seckel, 
 Nucl.\ Phys.\ B \textbf{283}, 681 (1987) {[}Erratum-ibid.\ B \textbf{296},
1034 (1988){]}. 
 

\bibitem{Gould:1987ju} A.~Gould, 
 Astrophys.\ J.\ \textbf{321}, 560 (1987). 
 

\bibitem{deBlok:2009sp} W.~J.~G.~de Blok, 
 Adv.\ Astron.\ \textbf{2010}, 789293 (2010) {[}arXiv:0910.3538
{[}astro-ph.CO{]}{]}. 
 

\bibitem{Moore:1994yx} B.~Moore, 
 Nature \textbf{370}, 629 (1994). 


\bibitem{Flores:1994gz} R.~A.~Flores and J.~R.~Primack, 
 Astrophys.\ J.\ \textbf{427}, L1 (1994) {[}astro-ph/9402004{]}.


\bibitem{Navarro:1996gj} J.~F.~Navarro, C.~S.~Frenk and S.~D.~M.~White,
 Astrophys.\ J.\ \textbf{490}, 493 (1997) {[}astro-ph/9611107{]}.
 

\bibitem{Spergel:1999mh} D.~N.~Spergel and P.~J.~Steinhardt,
 Phys.\ Rev.\ Lett.\ \textbf{84}, 3760 (2000) {[}astro-ph/9909386{]}.
 

\bibitem{Randall:2007ph} S.~W.~Randall, M.~Markevitch, D.~Clowe,
A.~H.~Gonzalez and M.~Bradac, 
 Astrophys.\ J.\ \textbf{679}, 1173 (2008) {[}arXiv:0704.0261 {[}astro-ph{]}{]}.


\bibitem{Rocha:2012jg} M.~Rocha, A.~H.~G.~Peter, J.~S.~Bullock,
M.~Kaplinghat, S.~Garrison-Kimmel, J.~Onorbe and L.~A.~Moustakas,
 Mon.\ Not.\ Roy.\ Astron.\ Soc.\ \textbf{430}, 81 (2013) {[}arXiv:1208.3025
{[}astro-ph.CO{]}{]}. 


\bibitem{Peter:2012jh} A.~H.~G.~Peter, M.~Rocha, J.~S.~Bullock
and M.~Kaplinghat, 
 arXiv:1208.3026 {[}astro-ph.CO{]}. 
 

\bibitem{Zavala:2012us} J.~Zavala, M.~Vogelsberger and M.~G.~Walker,
 Monthly Notices of the Royal Astronomical Society: Letters \textbf{431},
L20 (2013) {[}arXiv:1211.6426 {[}astro-ph.CO{]}{]}. 
 

\bibitem{Albuquerque:2013xna} I.~F.~M.~Albuquerque, C.~Prez de
Los Heros and D.~S.~Robertson, 
 JCAP \textbf{1402}, 047 (2014) {[}arXiv:1312.0797 {[}astro-ph.CO{]}{]}.


\bibitem{Aartsen:2012kia} M.~G.~Aartsen \textit{et al.} {[}IceCube
Collaboration{]}, 
 Phys.\ Rev.\ Lett.\ \textbf{110}, no. 13, 131302 (2013) {[}arXiv:1212.4097
{[}astro-ph.HE{]}{]}.

\bibitem{Zentner2009} A. R. Zentner, Phys. Rev. D \textbf{80}, 063501
(2009).

\bibitem{Cushman:2013zza} For a recent review on DM direct search,
see, for example, P.~Cushman, C.~Galbiati, D.~N.~McKinsey, H.~Robertson,
T.~M.~P.~Tait, D.~Bauer, A.~Borgland and B.~Cabrera \textit{et
al.}, 
 arXiv:1310.8327 {[}hep-ex{]}. 

\bibitem{Aartsen2014} M. G. Aartsen \emph{et al.} {[}IceCube-PINGU
Collaboration{]}, arXiv:1401.2046 {[}physics.ins-det{]}.

\bibitem{Feng:2014uja} J.~L.~Feng, S.~Ritz, J.~J.~Beatty, J.~Buckley,
D.~F.~Cowen, P.~Cushman, S.~Dodelson and C.~Galbiati \textit{et
al.}, 
 arXiv:1401.6085 {[}hep-ex{]}. 


\bibitem{Bertone2005}G. Bertone, D. Hooper and J. Silk, Phys. Rept.
\textbf{405}, 279 (2005).

\bibitem{Busoni2013}G. Busoni, A. De Simone and W. -C. Huang, JCAP
\textbf{1307}, 010 (2013) {[}arXiv:1305.1817 {[}hep- ph{]}{]}.

\bibitem{Chen2014}C.-S. Chen \emph{et al.}, JCAP \textbf{10}, 049
(2014) {[}arXiv:1408.5471 {[}hep-ph{]}{]}.

\bibitem{Blennow:2007tw} M.~Blennow, J.~Edsjo and T.~Ohlsson,
JCAP \textbf{0801}, 021 (2008). 


\bibitem{Aartsen:2012uu} M.~G.~Aartsen \textit{et al.} {[}IceCube
Collaboration{]}, 
Phys.\ Rev.\ Lett.\ \textbf{110}, 151105 (2013). 


\bibitem{Honda:2006qj} M.~Honda \textit{et al.}, 
Phys.\ Rev.\ D \textbf{75}, 043006 (2007). 


\bibitem{LUX2013} D. S. Akerib \emph{et al.} {[}LUX Collaboration{]},
Phys. Rev. Lett. \textbf{112}, 091303 (2014).\end{thebibliography}
\end{document}